\documentclass[%
 reprint,
 twocolumn,
superscriptaddress,
nofootinbib,
 amsmath,amssymb,
 aps,
 pra,
floatfix,
]{revtex4-2}

\usepackage{microtype}

\usepackage{algpseudocode}
\usepackage{booktabs}
\usepackage{graphicx,
            caption,
            subcaption,
            ragged2e,
            xcolor,
            xspace,
            textcomp,
            hyperref,
            scrextend, 
            placeins, 
            tabularx,
            }
\usepackage[nospace]{varioref}
\usepackage[capitalise]{cleveref} 

\DeclareCaptionType{algorithm}[ALG.]
\DeclareCaptionLabelSeparator{dot}{. }
\makeatletter
\def\justified{
	\let\\\@normalcr
	\@rightskip\z@skip \rightskip\@rightskip
	\leftskip\z@skip
	\parindent 0em\relax
	\setlength{\parfillskip}{0pt plus 1fil}}
\DeclareCaptionJustification{justified}{\justified}

\hypersetup{colorlinks=true, linkcolor=blue,citecolor=blue,urlcolor=blue}

\usepackage{scalerel}
\usepackage{tikz}
\usetikzlibrary{svg.path}
\definecolor{orcidlogocol}{HTML}{A6CE39}
\tikzset{
  orcidlogo/.pic={
    \fill[orcidlogocol] svg{M256,128c0,70.7-57.3,128-128,128C57.3,256,0,198.7,0,128C0,57.3,57.3,0,128,0C198.7,0,256,57.3,256,128z};
    \fill[white] svg{M86.3,186.2H70.9V79.1h15.4v48.4V186.2z}
                 svg{M108.9,79.1h41.6c39.6,0,57,28.3,57,53.6c0,27.5-21.5,53.6-56.8,53.6h-41.8V79.1z M124.3,172.4h24.5c34.9,0,42.9-26.5,42.9-39.7c0-21.5-13.7-39.7-43.7-39.7h-23.7V172.4z}
                 svg{M88.7,56.8c0,5.5-4.5,10.1-10.1,10.1c-5.6,0-10.1-4.6-10.1-10.1c0-5.6,4.5-10.1,10.1-10.1C84.2,46.7,88.7,51.3,88.7,56.8z};
  }
}
\newcommand\orcidicon[1]{\href{https://orcid.org/#1}{\mbox{\scalerel*{
\begin{tikzpicture}[yscale=-1,transform shape]
\pic{orcidlogo};
\end{tikzpicture}
}{|}}}}

\usepackage[ngerman, main=british]{babel}            
\useshorthands{"}

\makeatletter
\@ifpackageloaded{babel}%
    {%
        \addto\extrasamerican{%
            \languageshorthands{ngerman}%

			}%
		\addto\extrasenglish{%
            \languageshorthands{ngerman}}
        \addto\extrasbritish{%
            \languageshorthands{ngerman}}%
        \addto\extrasngerman{%
            }%
    }{\relax}
\makeatother

\begin{document}

\author{Maximilian Sohmen~\orcidicon{0000-0002-5043-2413}}
    \thanks{Correspondence should be addressed to\\ \url{maximilian.sohmen@i-med.ac.at}}
\affiliation{Institute for Biomedical Physics, Medical University of Innsbruck, 
6020~Innsbruck, Austria}

\author{Juan D. Muñoz-Bolaños~\orcidicon{0000-0003-2872-4126}}
\affiliation{Institute for Biomedical Physics, Medical University of Innsbruck, 
6020~Innsbruck, Austria}

\author{Pouya Rajaeipour~\orcidicon{0000-0003-2459-7545}}
\affiliation{Phaseform~GmbH, 
79110~Freiburg, Germany}

\author{Monika Ritsch-Marte~\orcidicon{0000-0002-5945-546X}}
\affiliation{Institute for Biomedical Physics, Medical University of Innsbruck, 
6020~Innsbruck, Austria}

\author{Çağlar Ataman~\orcidicon{0000-0002-6280-8465}}
\affiliation{Phaseform~GmbH, 
79110~Freiburg, Germany}
\affiliation{Microsystems for Biomedical Imaging Laboratory,  Department of Microsystems Engineering, University of Freiburg,
79110~Freiburg, Germany}

\author{Alexander Jesacher~\orcidicon{0000-0003-4285-9406}}
\affiliation{Institute for Biomedical Physics, Medical University of Innsbruck, 
6020~Innsbruck, Austria}

\title{Optofluidic adaptive optics in multi-photon microscopy} 

\begin{abstract}
Adaptive optics in combination with multi-photon techniques is a powerful approach to image deep into a specimen.
Remarkably, virtually all adaptive optics schemes today rely on wavefront modulators which are reflective, diffractive, or both.
This, however, can pose a severe limitation for applications.
Here, we present a fast and robust sensorless adaptive optics scheme adapted for transmissive wavefront modulators.
We study our scheme in numerical simulations and in experiments with a novel, optofluidic wavefront shaping device which is transmissive, refractive, polarisation-independent and broadband.
We demonstrate scatter correction of two"=photon"=excited fluorescence images of microbeads as well as brain cells and benchmark our device against a liquid-crystal spatial light modulator.
Our method and technology could open new routes for adaptive optics in scenarios where previously the restriction to reflective and diffractive devices may have staggered innovation and progress.
\end{abstract}

\date{\today}

\maketitle


\section{Introduction}

Optical microscopy is a potent and indispensable tool in many branches of science, in particular for biomedical research.
However, due to light scattering inside the sample, conventional light microscopy is typically limited to the sample's most superficial tens of micrometres.
Prominent approaches to lift this limitation and increase the imaging depth include multi-photon techniques~\cite{Denk1990} and adaptive optics (AO)~\cite{kubby2013adaptive, booth2014adaptive}.
For the present work, we rely on a combination of both, two-photon excited fluorescence (TPEF) microscopy as well as sensorless AO -- an approach that has been followed successfully in a variety of different ways in the recent past, by many groups worldwide~\cite{olivier2009dynamic, jesacher2009adaptive, Ji2010, gould2012adaptive, bourgenot20123d, tang2012superpenetration, papadopoulos2017scattering, hu2020universal}, including ours~\cite{may2021fast, may2021simultaneous, Sohmen:2022}.

AO relies on active shaping of a light field's wavefront to improve image quality.
The overwhelming majority of wavefront shaping devices on the market today --~such as deformable mirrors, liquid"=crystal"=on"=silicon spatial light modulators (LCoS-SLMs), or micro\-opto\-electro\-mecha\-ni\-cal systems (MOEMS)~\cite{Stewart:2007,Solgaard:2014} -- operates in \textit{reflection} rather than \textit{transmission} of a light field.
In addition, many of these devices (of the above, e.g., LCoS-SLMs and MOEMS) are \textit{diffractive} rather than \textit{reflective} optical elements.
While being reflective or diffractive is no fundamental problem in numerous cases, in others -- some of them very relevant for applications -- it can present a serious challenge.

A novel kind of transmissive, refractive wavefront modulator that may present a well-suited alternative for such scenarios is the deformable phase plate (DPP), a transparent multi-electrode optofluidic device recently developed by some of us~\cite{Rajaeipour:2019, Banerjee:2019, Rajaeipour:2020, Rajaeipour:2020_cascading, Rajaeipour:2021}.
The following three aspects may help to illustrate where the DPP can offer advantages compared to established wavefront modulators.

First, \textit{device integration}: 
compared to a reflective device,
integrating the DPP into a pre-existing (e.g., commercial) imaging system is much easier and does not require additional beam folding optics. 
Similarly, stacking several wavefront modulators in series (as, e.g., in woofer-tweeter arrangements~\cite{Rajaeipour:2020_cascading} or for multi-conjugate AO) is far more straightforward using transmissive elements.
Second, \textit{polarisation independence}: 
whereas, e.g., liquid-crystal SLMs can only shape light that is linearly polarised along a specific direction, the DPP can shape light of arbitrary polarisation.
This allows, on the one hand, to freely combine the DPP with polarisation-optical elements or, on the other hand, to shape unpolarised light such as the fluorescence from a microscopy sample.
Third, \textit{low chromatic dispersion}: 
whereas, by principle, pixelated holograms -- displayed, e.g., on a liquid-crystal SLM or a MOEMS -- are highly dispersive and typically only allow to shape wavelengths up to some tens of nanometres around a chosen design value, the DPP -- with a dispersion as low as less than 1~mm of fused silica~\cite{Rajaeipour:2021}~-- enables broadband operation.
Two immediate benefits are evident:
on the one hand, in applications involving short laser pulses, a drastic reduction of optical peak intensity due to pulse dispersion can be avoided.
On the other hand, in the imaging context, broadband operation offers, e.g., to address different fluorophore classes in parallel using several excitation wavelengths, or to route both excitation light and returning (even multi-photon-excited) fluorescence signal through the device on a shared path (e.g., when imaging back onto a pinhole as in confocal microscopy).

\bigskip

In our opinion, a practical AO approach that seeks to fully exploit the benefits of the DPP should therefore be guided by the following key criteria:
(i) \textit{simple integration} -- maintaining the possibility to easily insert the DPP into the optical path of a microscope;
(ii) \textit{high speed} -- carrying out a complete AO correction run should be fast (timescale seconds or less);
(iii) \textit{accuracy} -- measuring and compensating aberrations as accurately as possible;
(iv) \textit{robustness and user-friendliness} -- the corrections should converge reliably without requiring parameter tweaking by the user, ideally over a wide range of aberration severity~\cite{Sohmen:2022}.

This work is structured as follows. 
In Section~\ref{sec:approach},
we present our approach to meet the above criteria (i)--(iv), including a brief description of the DPP device, our wavefront sensing strategy, and sketches of the optical layout as well as our AO algorithm. 
In Section~\ref{sec:numerics}, we compare different algorithm variants in numerical simulations.
In Section~\ref{sec:experiments}, we experimentally demonstrate correction of strong aberrations through combination of our sensorless AO algorithm with DPP wavefront shaping. 
To this end, we present TPEF microscope images of two kinds of samples, standardised fluorescent beads as well as fluorescence-labelled brain cells.
Finally, we discuss our main findings (Section~\ref{sec:discussion}), give a brief Summary and Outlook (Section~\ref{sec:summaryoutlook}), and provide information on our Materials and Methods (Section~\ref{sec:materialsmethods}).


\section{General approach}  
\label{sec:approach}

Before we proceed to our numerical simulations and experimental results, we will sketch our general approach and the design considerations that have guided us in our development process.


\subsection{The deformable phase plate (DPP)}
\label{sec:dpp}

\begin{figure}[tb] 
    \centering
    \includegraphics[width = \columnwidth]{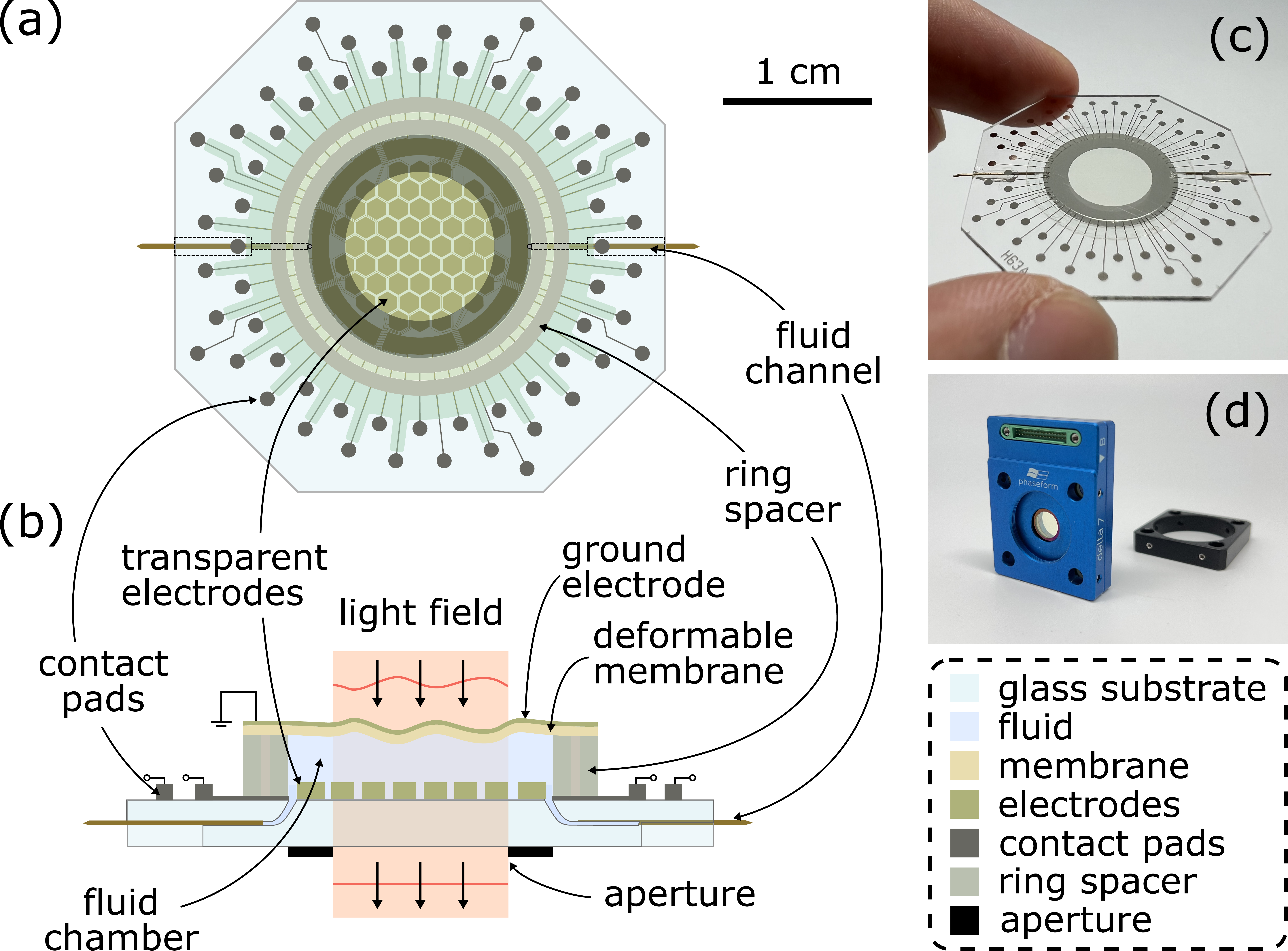}
    \caption{\textbf{The deformable phase plate (DPP).} (a)~Top-view drawing. 
    Each of the transparent electrodes in the centre region can be set to an individual electric potential through connected contact pads.
    (b)~Cross-sectional drawing and principle of operation.
    A voltage between an array electrode and the grounded membrane results in a membrane deformation. 
    The locally varying optical path length leads to a wavefront modulation of an oncoming light field.
    (c)~Photograph of the DPP naked and (d) readily assembled in its 30-mm cage-compatible housing.
    }
    \label{fig:DPPphoto}
\end{figure}

The DPP is a transparent optofluidic device designed for wavefront modulation of a light field \textit{in transmission}~\cite{Rajaeipour:2019, Banerjee:2019, Rajaeipour:2020, Rajaeipour:2020_cascading, Rajaeipour:2021}.
The DPP's fluid chamber is formed by a micro-machined ring spacer, placed on a glass substrate and spanned by a deformable polymeric membrane (see Fig.\,\ref{fig:DPPphoto}).
Channels in the glass substrate through which the chamber is filled with liquid are sealed after manufacture.
On the glass substrate, contact pads in the periphery individually connect to 63 transparent electrodes in and around the central aperture.
Application of a voltage between any of these electrodes and the grounded, conductive membrane gives rise to an electrostatic force which deforms the membrane. 
If, consequently, the fluid chamber's local thickness changes from $\ell$ to $\ell' = \ell + \Delta\ell$, the corresponding phase shift for a transmitted light field of wavelength $\lambda$ is $\Delta\varphi = 2\pi (\Delta\ell/\lambda) \Delta n$, where $\Delta n$ is the according difference in refractive index between the fluid and air.
While electrostatic actuation alone would only enable a unidirectional pulling of the membrane, hydro-mechanical coupling by the liquid inside the sealed chamber establishes a bidirectional push-pull mechanism, where a positive displacement in one membrane portion leads to a negative displacement in the other portions, and vice versa. 

The DPP action is independent of polarisation, free from diffractive losses, nearly non-dispersive, gravity-neutral (see Methods), and shows no observable hysteresis, allowing for operation with open-loop control.
For wavelengths between 400 and 2200~nm, the present DDP (Delta~7, Phaseform GmbH, Freiburg i.\,Br., Germany) shows \mbox{$>85\,\%$} transmission of power, limited primarily by Fresnel reflection at uncoated surfaces; only below 400~nm absorption by the polymeric membrane becomes appreciable.
The maximum stroke of the DPP is highest for low-order Zernike modes, e.g., about $\pm 4$~µm (peak-valley) optical path difference for a defocus aberration at 632~nm wavelength.
The rise time (10\,\% to 90\,\% of set value) of the DPP is about 50~ms, limited by fluid flow.


\subsection{Wavefront sensing strategy}
\label{sec:wavefrontsensing}

The idea of sensorless AO is to conduct many cycles of applying test modes to the wavefront modulator while measuring their effect on the target signal (e.g., fluorescence intensity) and to construct a wavefront correction pattern from this information~\cite{kubby2013adaptive,booth2014adaptive}.
In our case, however, faced with the comparatively long switching time of the DPP, it is desirable to follow a wavefront sensing strategy where taking fast measurements of individual modes is \textit{decoupled} from the limited wavefront modulator speed.

Focus Scanning Holographic Aberration Probing (F"~SHARP), as introduced by Papadopoulos et al.~\cite{papadopoulos2017scattering}, was one of the first methods to employ such a decoupling.
The basic principle of F"~SHARP is to split the laser beam for fluorescence excitation in an interferometer and to vary the angle between the two interferometer arms -- a stronger, static reference beam and a weaker, scanned probe beam -- using a fast piezo mirror.
The recorded interferogram then directly reveals the aberrations accumulated along the optical path to the sample plane, which can in turn be compensated by a wavefront modulator. 
The wavefront modulator itself is kept constant during the (fast) piezo scanning operation and only has to be switched \textit{once} per algorithm iteration (after the aberration field has been determined), wherefore the wavefront modulator switching time is in general not too limiting for the total algorithm speed.
Besides this speed advantage, F"~SHARP exhibits a very \textit{robust} operation:
in contrast, e.g., to modal wavefront sensing techniques \cite{olivier2009dynamic,jesacher2009adaptive}, the F"=SHARP algorithm converges reliably even if the aberrations are comparatively strong~\cite{papadopoulos2017scattering}.

To-date, F"~SHARP-type methods have been used exclusively in combination with LCoS-SLMs~\cite{papadopoulos2017scattering, Cui2011, papadopoulos2020dynamic, qin2022deep}, i.e., wavefront modulators that are \textit{reflective} and \textit{diffractive}.
Here, we present and test a novel variant fit for use in combination with \textit{transmissive} (and, as in case of the DPP, \textit{refractive}) wavefront modulators.


\subsection{Optical layout}
\label{sec:setup}

\begin{figure}[tb] 
    \centering
    \includegraphics[width = 0.9\columnwidth]{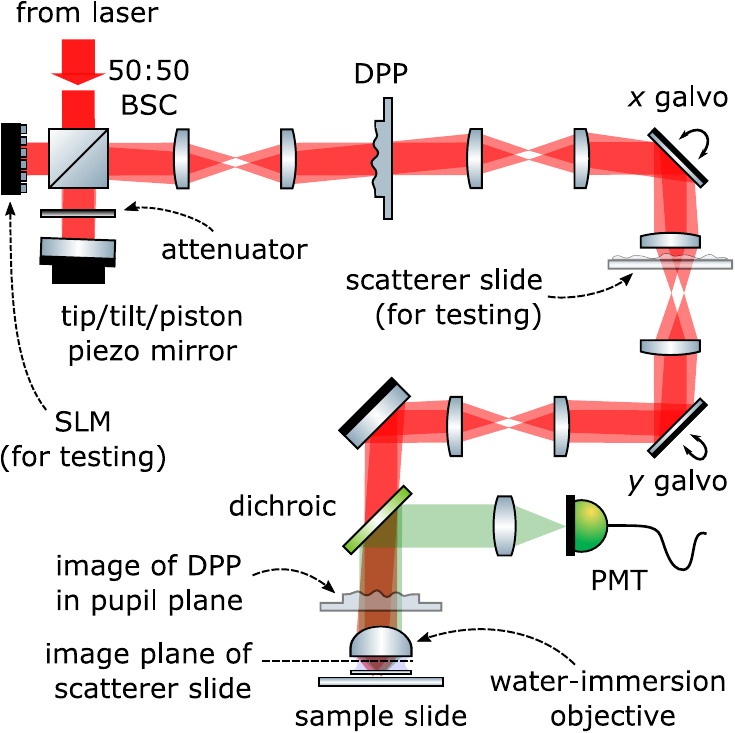}
    \caption{\textbf{Sketch of the optical layout.} BSC = (non-polarising) beam splitter cube, DPP = deformable phase plate, PMT = photomultiplier tube, SLM = spatial light modulator.}
    \label{fig:setup}
\end{figure}

A schematic of our optical setup is presented in Fig.\,\ref{fig:setup}.
This setup is based on the original F"~SHARP implementation~\cite{papadopoulos2017scattering}, yet with two crucial modifications.
First, instead of a Mach-Zehnder, our setup features a Michelson interferometer, bearing the advantage that distances which are not common path between the two arms can be kept as short as possible, helping to minimise relative vibrations, drifts, and alignment maintenance.
Second, instead of shaping \textit{only} the static reference beam, in our DPP version \textit{both}, static reference and scanned probe beam are wavefront-shaped.

These two modifications greatly simplify integrating F"~SHARP into pre-existing optical setups: the Michelson interferometer, on the one hand, can be designed as a highly stable and compact add-on module in the early excitation path; the DPP, on the other hand, being transmissive, can be inserted almost everywhere along the downstream optical path.
In our case, the DPP is located close to a pupil-conjugate plane (within some millimetres accuracy).


\subsection{Algorithm outline}
\label{sec:algorithm}

\begin{figure}[b] 
    \centering
    \includegraphics[width = \columnwidth]{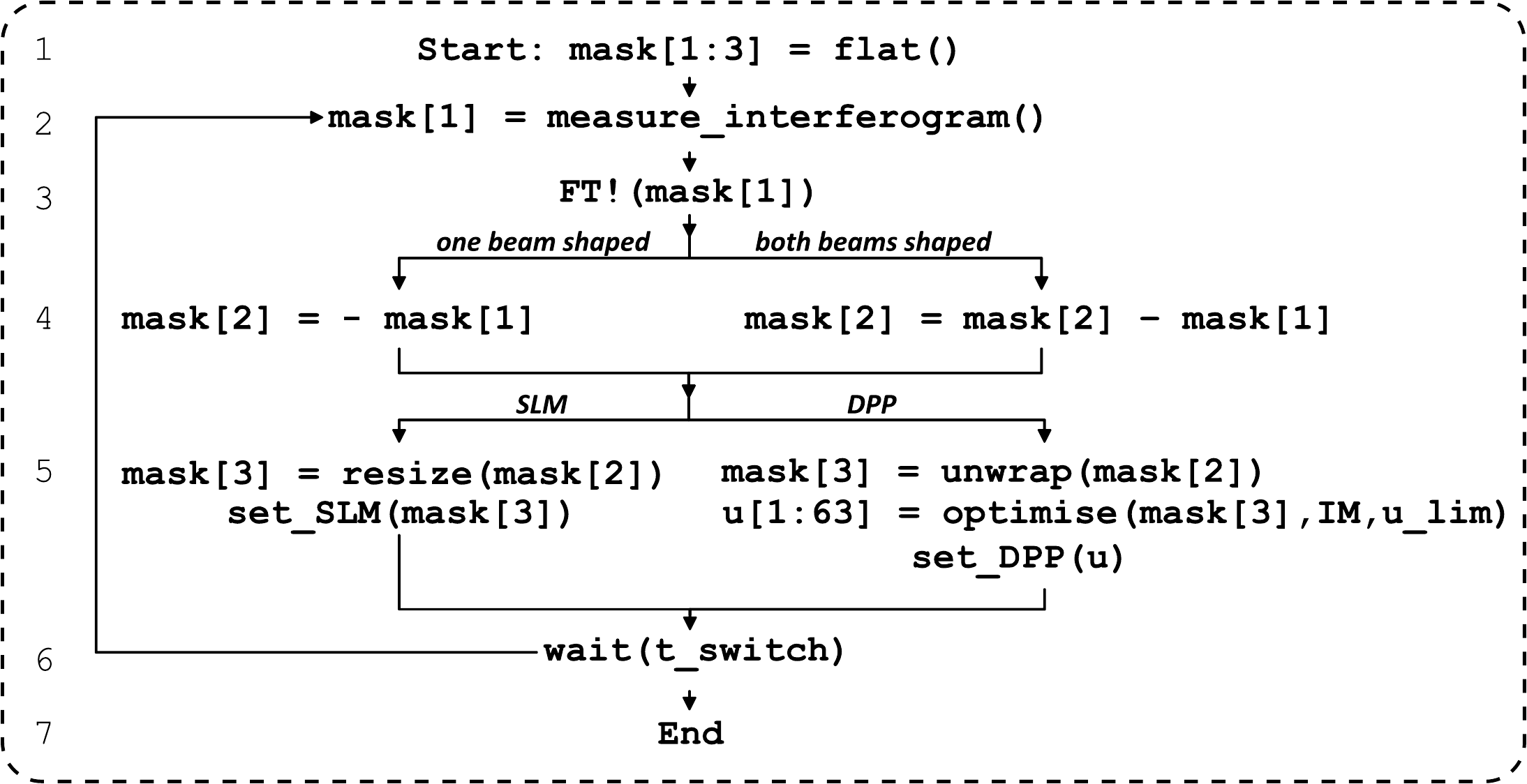}
    \caption{\textbf{Diagrammatic algorithm outline}. Overview of critical steps shared and differing between the original and our F"~SHARP variant. 
    Functions with names ending with an exclamation mark modify their argument. 
    \texttt{FT!($\ldots$)}: Fourier transform;
    \texttt{optimise($\ldots$)}: find the voltages that optimally reproduce a target phase mask (e.g., in Zernike basis) at given constraints~\cite{Rajaeipour:2019}. \texttt{IM} denotes the influence matrix, \texttt{u\_lim} the voltage limit of the DPP electrodes (see main text).}
    \label{fig:algorithm}
\end{figure}

Figure~\ref{fig:algorithm} provides a diagrammatic overview over important algorithm steps shared and differing between the original and our variant of F"~SHARP.

In Step~1, the phase correction mask is initialised flat.
In Step~2, by phase-stepping the piezo in the probe arm at a range of different tip/tilt angles and measuring the TPEF intensity generated in the object plane, we interferometrically obtain an estimate of the aberrations in the excitation path. 
In Step~3, the field obtained through interferometry is propagated numerically to the wavefront modulator location (in our case, where the DPP is pupil-conjugate, this corresponds to a Fourier transform).
In Step~4, the old phase mask (from a previous iteration) is either \textit{replaced} by the negative new mask in case \textit{only} the static beam is shaped (as for F"~SHARP with SLM; abbreviated 1-B), or the new mask is \textit{subtracted} from the old mask in case \textit{both} beams are shaped (as with DPP; abbreviated 2-B). 
In Step~5, the target phase mask is prepared for display.
This includes minor numerical adjustments such as, e.g., rotations and flips to account for the relative orientation between optical elements (e.g., the wavefront modulator and the piezo scan axes). 
For diffractive wavefront modulators such as an SLM, the target phase mask needs to be resized from the number of measured (piezo scan) pixels to the number of displayed (SLM) pixels.
For refractive wavefront modulators such as the DPP, the target phase mask needs to be phase unwrapped.
Subsequently, we decompose the target phase mask into Zernike modes and determine the DPP electrode voltages which lead to the closest possible reproduction of this mode combination, following the optimisation procedure of Ref.~\cite{Rajaeipour:2019}.
After sending the update command to the wavefront modulator, we pause in Step~6 until the device has switched.
Afterwards, the algorithm either ends or jumps back to Step~2 for another iteration.


\section{Results}
\label{sec:results}

\FloatBarrier

\subsection{Numerical simulations}
\label{sec:numerics}

\begin{table}[tb]
    \caption{\textbf{F"~SHARP variants covered in this work.}
    }
    \centering
    \begin{tabularx}{\columnwidth}{m{35mm} X X X X X}
        \toprule
        \textbf{Variant} & \textbf{1-B} & \textbf{2-B} & \textbf{2-BI}\\
        \midrule
        number of shaped beams & 1 & 2 & 2\\
        phase unwrapping & $\times$ & $\times$ & \checkmark\\
        influence matrix & $\times$ & $\times$ & \checkmark\\
        Zernike decomposition & $\times$ & $\times$ & \checkmark\\
        \bottomrule
    \end{tabularx}
    \label{tab:methods}
\end{table}

\begin{figure}[tbh] 
    \centering
    \includegraphics[width = \columnwidth]{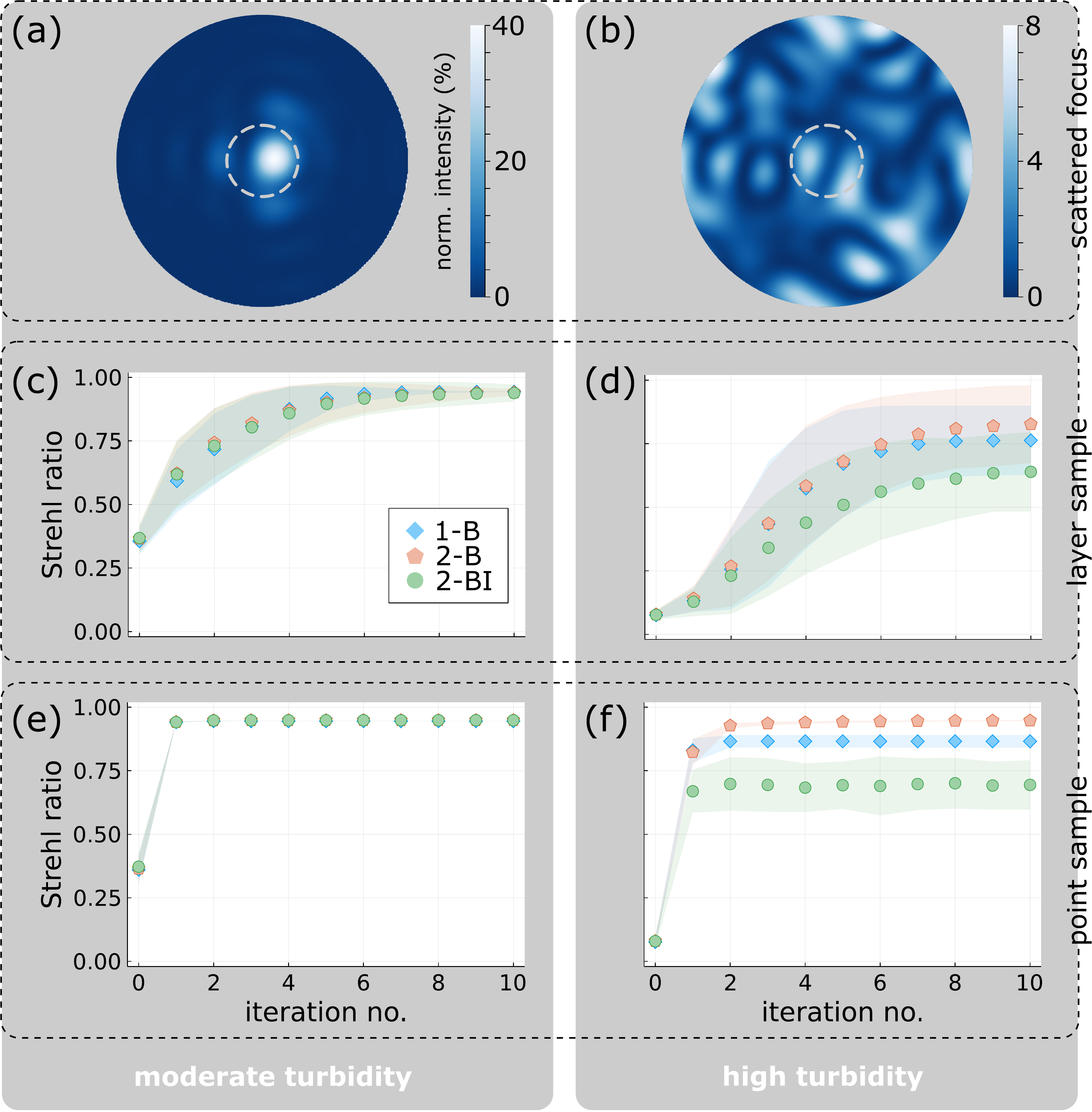}
    \caption{\textbf{Simulated algorithm progress.}
    Comparison of simulation results for moderate ($L/l_s = 1, \sigma/d_0 = 1$; \textit{left}) and high turbidity ($L/l_s = 3, \sigma/d_0 = 3$; \textit{right}). 
    (a,\,b)~Examples of the focal-plane intensity distributions \textit{before} AO correction, for scatter masks at the respective turbidity levels.
    The colour scale represents normalised intensity, such that the respective maximum value equals the Strehl ratio.
    For comparison, the grey dashed circles give the Airy disk diameter (\O$/2{}=1.22 \, d_0$) of a perfect focus.
    (c,\,d)~Improvement of Strehl ratio with algorithm iterations for a 2D fluorescent layer sample.
    The different methods plotted are F"~SHARP correcting one (1-B) or two beams, with (2-BI) and without (2-B) considering the influence matrix and voltage limits of the DPP.
    Iteration~0 corresponds to the value before correction.
    Markers give the mean, colour-shaded areas the standard deviation over 100 simulation runs with different random scatter masks at the respective level of turbidity.
    (e,\,f)~Corresponding Strehl ratio improvements assuming a point-like fluorescent sample.
    }
    \label{fig:simu_scatter}
\end{figure}

We have performed numerical simulations of our AO variants to accompany the experiments presented in the following Section~\ref{sec:experiments}.
Our numerical model, whose general outline is sketched in Fig.\,\ref{fig:algorithm}, is based on the code provided with Ref.~\cite{may2021fast}. 
In brief, we consider an excitation light field -- disturbed by a phase-only scatter mask -- which illuminates a fluorescent sample in the focal plane. 
Our random scatter masks are located in the pupil, obtained following the recipe described in Ref.~\cite{Sohmen:2022}, and characterised by two parameters, a `thickness', $L$, and the width of their spatial-frequency content, $\sigma$.
We state $L$ in units of the photon scattering mean free path, $l_s$; note that in many animal or human tissues photons are scattered predominantly into the forward direction, such that it typically takes around ten consecutive scattering events until a photon's travelling direction becomes fully randomised (i.e., to enter the diffusive radiation transport regime)~\cite{Cheong:1990,Ntziachristos2010}.
The spatial frequency distribution of the scatterer is assumed Gaussian, centred about zero, and its standard deviation, $\sigma$, is expressed in units of the spatial resolution, $d_0=\lambda/(2\,\text{NA})$, where $\text{NA}$ denotes the numerical aperture.

Our simulations follow the algorithmic procedure presented in Section~\ref{sec:algorithm}.
For each iteration, we consider $N_\text S = 60$ different piezo mirror tip/tilt angles (in $k$-space, these can be imagined as the elements within the incircle of a $10{\times}10$ square grid), which is close to the number of degrees of freedom of the DPP, 63.
We mainly present simulation results for a maximally unstructured sample -- a homogeneous 2D fluorescent layer --, but complement this with examples for a maximally structured, point-like fluorescent sample.
Note that, generally, the layer sample is `tougher' for indirect wavefront sensing than the point-like sample. 
Furthermore, in this work we aim primarily at high"=signal"=level scenarios, where we can safely neglect the effect of photon shot noise in our simulations.
We numerically compare three different algorithm variants:
F"~SHARP correcting one (1-B) or two beams, with (2-BI) and without (2-B) considering the influence matrix of the DPP.
This influence matrix relates a target phase pattern to the voltages that need to be applied to the 63 electrodes, thus incorporating the DPP's limited voltage (i.e., stroke) and spatial modulation frequency~\cite{Rajaeipour:2019}.
For the current DPP driver software, the required format of the target phase pattern was a vector of Zernike mode amplitudes of length 91 (i.e., up to 11th radial order).
The defining features of all variants are summarised in Table~\ref{tab:methods}.

Figures~\ref{fig:simu_scatter}\,(a) and (b) illustrate the effect of scatter masks at moderate ($L/l_s = 1, \sigma/d_0 = 1$) and high turbidity ($L/l_s = 3, \sigma/d_0 = 3$), respectively.
Compared to a perfect, Airy-shaped focal spot (grey dashed circle), optical power is spread out over the focal plane.
It is intuitively clear that scanning such a distorted `focus' over a sample will produce a worse image than a perfect focus.
Quantitatively, the deviation from a perfect focus is captured by a reduced Strehl ratio.
Starting from distorted `foci' like in Figs~\ref{fig:simu_scatter}\,(a,\,b), successive iterations of our correction algorithm variants can bring the Strehl ratio back close to unity, restoring an almost perfect focus.
This is plotted in Figs~\ref{fig:simu_scatter}\,(c,\,d) for the layer sample.
Whereas at moderate turbidity [Fig.\,\ref{fig:simu_scatter}\,(c)] all three variants exhibit a very similar performance, reaching a final Strehl ratio of about 0.95, at high turbidity [Fig.\,\ref{fig:simu_scatter}\,(d)] variants 2-B and 1-B [both 0.8(2)] perform slightly better than variant 2-BI [0.6(2)]. 
Here and throughout this work, a character in parentheses (if present) gives the one-$\sigma$ standard deviation of the last digit.
Figures~\ref{fig:simu_scatter}\,(e,\,f) show the corresponding performances for the point-like sample, where we observe that the algorithms reach their final value much quicker (not later than after the second iteration). 
Again, at moderate turbidity, all methods perform similarly, reaching a final Strehl ratio of about 0.95, whereas at high turbidity variants 2-B [0.947(2)] and 1-B [0.87(3)] perform better than variant 2-BI [0.7(1)].

\begin{figure}[tbh] 
    \centering
    \includegraphics[width = \columnwidth]{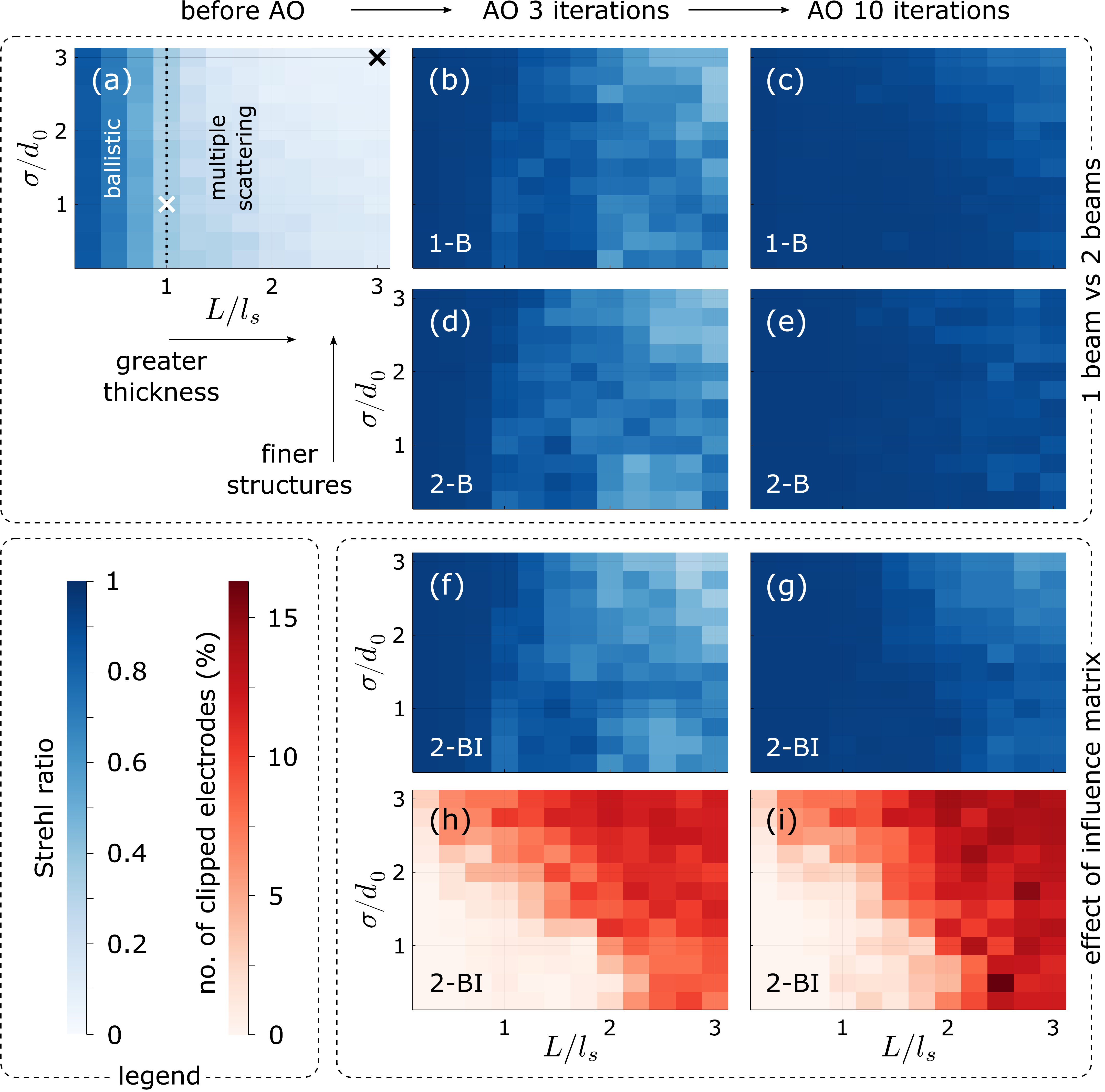}
    \caption{\textbf{Survey of simulated algorithm performance with respect to scatterer characteristics.} 
    Strehl ratios (blue colour scale) before correction (left column), after the 3rd (centre column) and after the 10th (right column) algorithm iteration for a layer sample, as a function of scatterer thickness, $L/l_s$, and spectral width, $\sigma/d_0$.
    Rows~1--3 correspond to the F"~SHARP variants 1-B, 2-B, and 2-BI, respectively (see main text).
    In subfigure~(a), the black dotted line marks the crossover from the ballistic to the multiple-scattering photon travel regime~\cite{Ntziachristos2010}, and the white and black cross correspond to the turbidity levels of the left and right grey box in Fig.\,\ref{fig:simu_scatter} \vpageref{fig:simu_scatter}, respectively.
    (h,\,i)~Row~4 (red colour scale) shows how many of the DPP electrodes have reached their voltage limit (`clip'), in relation to the total number of 63 electrodes.
    All data shown are the mean over 15 simulation runs with different random scatter masks.
    }
    \label{fig:simu_heatmaps}
\end{figure}

To illustrate how the algorithm performance depends on the scatterer characteristics (thickness $L$ and spectral width $\sigma$), we present respective overview simulations for a layer sample in  Fig.\,\ref{fig:simu_heatmaps}.
We compare the Strehl ratios before AO (a), reached after the 3rd (b,\,d,\,f) and after the 10th (c,\,e,\,g) algorithm iteration by variants 1-B, 2-B, and 2-BI, respectively.
As marked by the black dotted line in Fig.\,\ref{fig:simu_heatmaps}\,(a), the turbidity range covered by our analysis extends substantially beyond the ballistic ($L<l_s$) and into the multiple-scattering ($L>l_s$) photon travel regime~\cite{Ntziachristos2010}, where `traditional' sensorless AO methods (such as, e.g., the $3N$-algorithm~\cite{Booth2002_pnas}) typically fail.
Figures~\ref{fig:simu_heatmaps}\,(b,\,c) and (d,\,e) in general look quite similar and do not hint at stark systematic differences between the 1-B and the 2-B method with respect to the scatterer characteristics.
Comparing them to the 2-BI method [Figs~\ref{fig:simu_heatmaps}\,(f,\,g)], we see that taking the DPP influence matrix into account leads to a slightly lower performance, especially at high $L$ and/or $\sigma$.
This is plausible, since in general, the higher $L$ and $\sigma$, the higher will be the stroke (hence, voltage) of DPP electrodes required to display the unwrapped correction phase mask. 
The more of the 63 DPP electrodes clip at the internal voltage limit of 400~V, the less accurate becomes the compensation pattern, staggering the DPP correction performance at high turbidity.
The percentage of electrodes clipping at their voltage limit is shown in Figs~\ref{fig:simu_heatmaps}\,(h,\,i).

\subsection{Experiments}
\label{sec:experiments}

We analysed the performance of our new AO scatter correction approaches by experimental fluorescence imaging using a home-built two-photon microscope and the AO module outlined in Section~\ref{sec:setup}.
We compared the performance of the DPP and its adapted F"~SHARP variant (2-BI) to an LCoS-SLM (HSP1920-500-1200-HSP8, 1920$\times$1152 pixels, pixel size 9.2~µm, 
Meadowlark Optics, Inc., Frederick, CO, USA) with original F"~SHARP (1-B), cf.~Fig.\,\ref{fig:setup}. 
We investigated the performance of the AO schemes using two different kinds of samples, fluorescent beads (Section~\ref{sec:beads}) and brain cells labelled by green fluorescent protein (GFP, Section~\ref{sec:brain}).

We are aware that due to the big difference in the number of degrees of freedom between the DPP and the SLM a direct comparison is not strictly `fair'; our results should therefore be considered as benchmark tests of a new technology (which will be further developed in the future) against an established gold standard.


\subsubsection{Fluorescent beads}
\label{sec:beads}

\begin{figure*}[tb] 
    \centering
    \includegraphics[width = 0.75\textwidth]{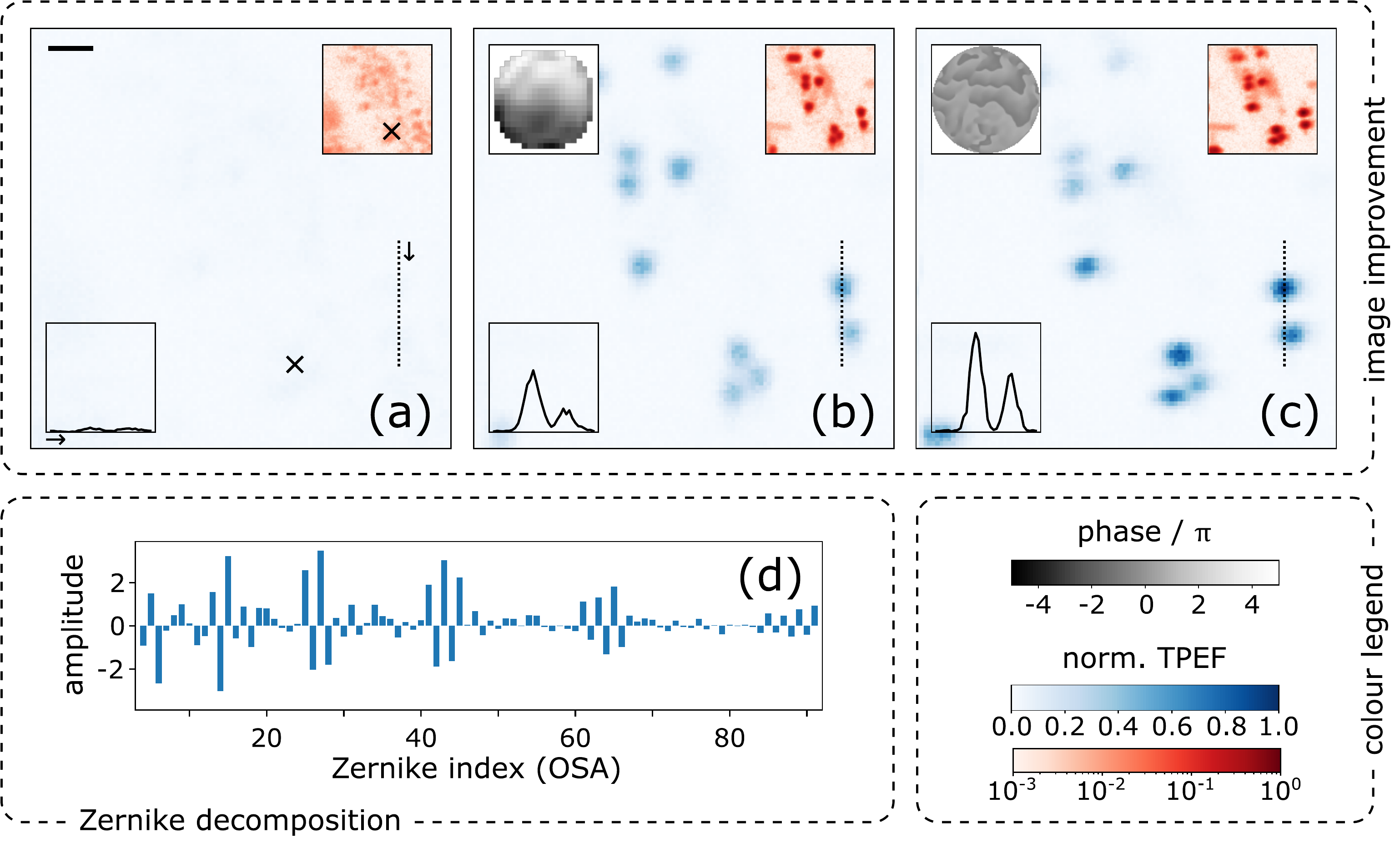}
    \caption{\textbf{Imaging fluorescent beads.} 
    Two-photon images of a sample of fluorescent beads (\O~0.5~µm), aberrated by a nail-polish slide (cf.~Fig.\,\ref{fig:setup}). 
    The colour scales are common for all subfigures~(a--c) and two-photon excited fluorescence (TPEF) intensities are normalised to the highest occurring value [\textit{here}: in (c)].
    (a)~The beads without AO. 
    The main panel gives the recorded TPEF intensity in linear (blue) colour scale. 
    The upper right inset is the same image in logarithmic (red) colour scale, the line plot shows an intensity cut along the dotted line in linear scale, and the top-left black scale bar corresponds to 1~µm. 
    The black cross ($\times$) marks the pixel selected for signal collection in subsequent AO trials.
    (b--c)~The same image window after application of our AO schemes.
    (b)~F"~SHARP using the DPP (blank SLM, 2-BI); the upper left inset shows the correction phase mask.
    (c)~F"~SHARP using the SLM (blank DPP, 1-B).
    (d)~The amplitudes of fitted Zernike modes (excluding piston, tip, tilt; in OSA single-indexing scheme~\cite{Thibos:2000}) corresponding to the correction of subfigure~(b); see main text.}
    \label{fig:beads}
\end{figure*}

In a first series of measurements, we tested our methods by imaging of 0.5-µm diameter fluorescent beads (TetraSpeck~T$7284$, Thermo Fisher Scientific, Inc., Waltham, MA, USA) dryed onto a glass slide, embedded in mounting medium and covered by a glass slip.
To introduce moderate aberrations, we placed a (nail-polish) scatterer slide in the light path (see Fig.\,\ref{fig:setup} and Methods).
To disentangle different effects, we  recorded sets of three different consecutive measurements. 
Figure~\ref{fig:beads} shows a representative example of such a measurement set, with all subfigures (a--c) sharing the same colour scales and TPEF intensities normalised to the highest value occurring in the set.

Figure~\ref{fig:beads}\,(a) shows a TPEF image of the beads (aberrated by effect of the scatterer slide) \textit{before} application of corrective AO, the length of the scale bar corresponding to 1~µm in the object plane. 
The main panel shows the recorded normalised TPEF signal in linear (blue) colour scale, where barely any structure is visible. 
Also the intensity cut along the dotted line, plotted in the lower left inset in linear scale, appears essentially flat.
Only in logarithmic colour scale (upper right inset; red colour scale), some diffuse patches are discernible and hint at an underlying structure. 

In contrast, Figs~\ref{fig:beads}\,(b,\,c) present the identical image window after applying the DPP and the SLM AO scheme, respectively, each running for 3~iterations and scanning over $N_\text S = 20{\times}20$ focal points using the piezo mirror.
The exact choice of $N_\text S$ is largely arbitrary and uncritical for our conclusions.
Note, however, that while in our simulations (where $N_\text S < 91$) the required switch from a (square) pixel-by-pixel to a Zernike basis is without consequences, in the experiments (where $N_\text S > 91$) both the Zernike decomposition of the target phase pattern as well as the restriction to 63 electrodes have a low-pass filtering effect and lead to an expected, systematic disadvantage of the DPP compared to the SLM correction performance.

Figure~\ref{fig:beads}\,(b) was recorded after running F"~SHARP (variant 2-BI), using solely the DPP for phase shaping (i.e., leaving the SLM constantly blank).
As described above, the procedure in every algorithm iteration involved interferometrical retrieval of a corrective phase mask (greyscale inset), unwrapping, decomposition into Zernike modes, determination of the optimum electrode voltages using the DPP influence matrix, and feeding the 63 control voltage values to the DPP.
Compared to Fig.\,\ref{fig:beads}\,(a), a striking improvement in terms of both signal level as well as structure clarity is observable in Fig.\,\ref{fig:beads}\,(b).
In the main panel as well as in the log-scale inset several distinct beads are clearly visible.
The total signal enhancement, by a factor of $\eta \sim 9$ (see Methods), is particularly evident in the intensity cut.
The retrieved phase mask leading to this improvement is shown in the upper left inset of Fig.\,\ref{fig:beads}\,(b); the magnitudes of fitted Zernike modes (normalised according to Ref.~\cite{Thibos:2000}) passed on for voltage optimisation are shown in the bar chart, Fig.\,\ref{fig:beads}\,(d).

In Fig.\,\ref{fig:beads}\,(c), we show the bead image after our benchmark AO scheme, i.e., running F"~SHARP using solely the SLM (constant, flat DPP; variant 1-B) and displaying a correction mask at the native SLM resolution (in the current case, using a display patch of 950${\times}$950 pixels).
Compared to Fig.\,\ref{fig:beads}\,(a), the signal enhancement in Fig.\,\ref{fig:beads}\,(c) is by a factor of $\eta \sim 14$. 

On the one hand, it seems plausible that the lower signal enhancement observed for the DPP compared to the SLM is, at least in part, caused by its lower number of degrees of freedom.
On the other hand, contrasting Figs~\ref{fig:beads}\,(b) and (c), it appears that the DPP correction is more `homogeneous' over the field of view. 
This is probably the downside of the SLM's higher number of degrees of freedom:
higher spatial frequency components will enhance the signal \textit{inside} the isoplanatic patch, but at the same time reduce the size of the isoplanatic patch and -- since the scatterer is not located in the pupil plane -- decrease the signal \textit{outside} the isoplanatic patch.
Therefore, if for an imaging application the size of the isoplanatic patch is of high importance, it can be advantageous to perform AO restricted to low spatial frequencies.

\bigskip

When starting from an aberrated (uncorrected) initial image, it is typically not clear to the user a priori which pixel should be picked for signal optimisation [see, e.g., the cross in Fig.\,\ref{fig:beads}\,(a)].
For typical samples (such as the beads or brain tissue studied in this work), however, the performance of an AO algorithm will depend on this choice, since in general between different pixels determining factors like aberrations or signal-to-noise ratio are not identical.
A practical measure of how much the performance of our algorithms depends on the particular choice of the target pixel is the relative spread of enhancement, $\langle\sigma_\eta / \eta\rangle$ (see Methods for definition).
We find $\langle\sigma_\eta / \eta\rangle \sim 50\,\%$ using the DPP and $\langle\sigma_\eta / \eta\rangle \sim 40\,\%$ using the SLM.


\subsubsection{Brain cells}
\label{sec:brain}

\begin{figure*}[tb] 
    \centering
    \includegraphics[width = 0.75\textwidth]{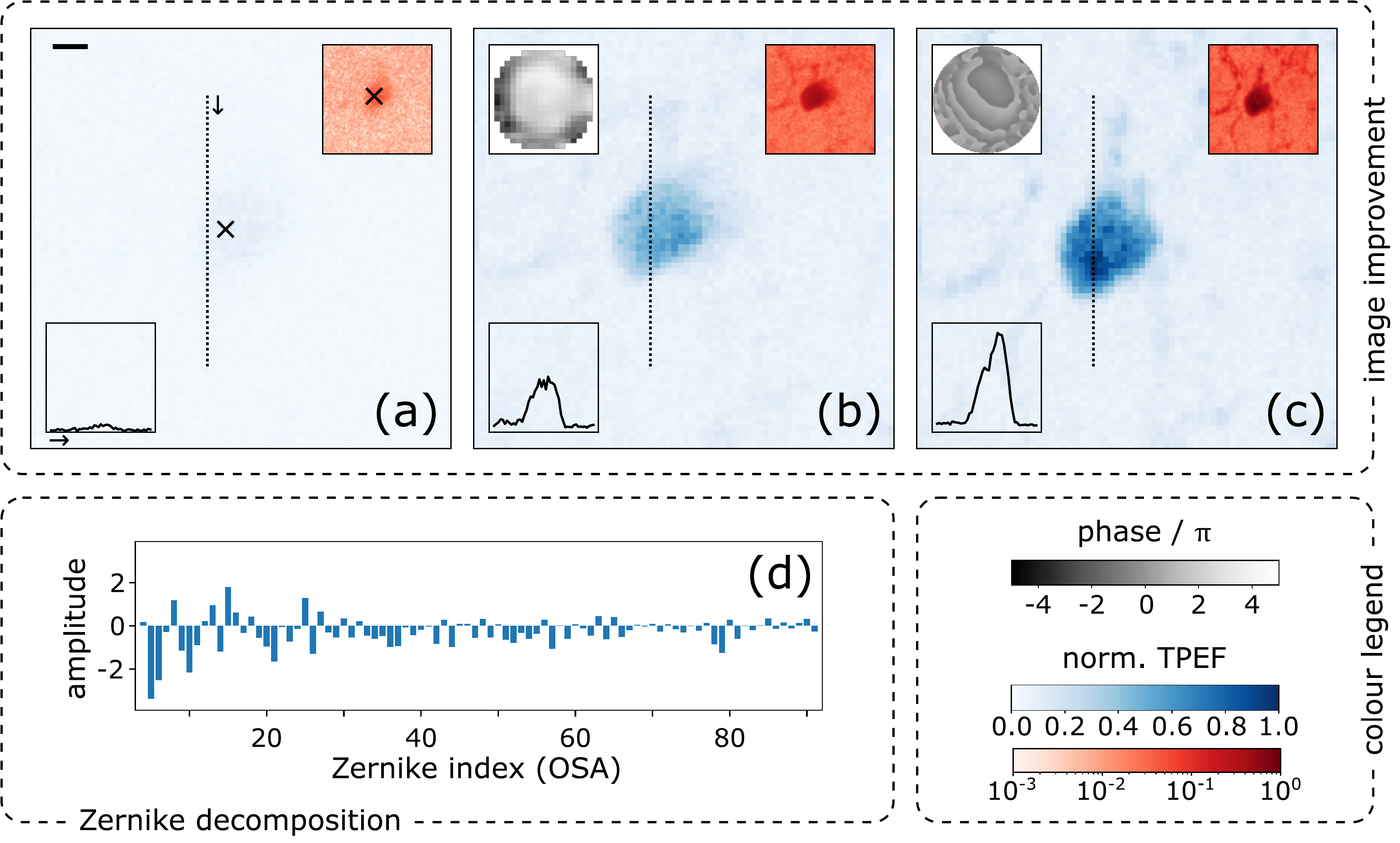}
    \caption{\textbf{Imaging into brain tissue.} 
    Two-photon fluorescence images of a microglia cell expressing GFP, located 0.2~mm deep inside fixed mouse brain tissue (\textit{no} scatterer slide).
    Correction procedures and presentation of data are analogous to Fig.\,\ref{fig:beads}; the scale bar corresponds to 1~µm.
    (a)~The cell aberrated by light scattering in the surrounding tissue, before correction. 
    (b)~F"~SHARP (2-BI) using the DPP.
    (c)~F"~SHARP (1-B) using the SLM.
    (d)~fitted Zernike mode amplitudes for the correction of subfigure~(b).}
    \label{fig:brain}
\end{figure*}

In a second series of measurements, we have tested our variants by imaging of microglia in coronal slices of fixed mouse brain (see Methods).
We imaged cells expressing GFP at a depth of about 0.2~mm, such that the light on the way in and out was aberrated by propagation through brain tissue (without scatterer slide).
For ease of comparison, the brain image data (Fig.\,\ref{fig:brain}) were taken in an identical procedure and are presented in analogous form to the bead data (Fig.\,\ref{fig:beads}) of the previous Section~\ref{sec:beads}.

Figure~\ref{fig:brain}\,(a) shows a fluorescence image of the glia cell aberrated by light scattering in its surrounding tissue, before application of AO. 
Whereas, again, neither in the linear scale of normalised TPEF intensity (main panel, blue) nor in the intensity cut clear structure is visible, the log-scale (red) inset reveals faint hints of the presence of a cell body.
Figures~\ref{fig:brain}\,(b,\,c) show the same cell after 3~iterations of our respective AO correction methods, again sampling 20${\times}$20 points around the focus of the strong beam using the piezo scanner.

Figure~\ref{fig:brain}\,(b) shows an image after running F"~SHARP (variant 2-BI) using solely the DPP.
A clear improvement in signal level as well as in structure clarity is found: in the linear-scale main panel as well as in the intensity cut we observe an enhancement of total signal by a factor of $\eta \sim 5$; especially in the log-scale inset, besides the pronounced cell soma, fine cellular processes start to become visible. 

Figure~\ref{fig:brain}\,(c) shows the image corrected using our SLM benchmark scheme (F"~SHARP variant 1-B).
As for the beads, with this method we obtain the visually clearest image with the highest signal enhancement of  $\eta \sim 8$.

\bigskip

For convenience of comparison, important parameters and results of the measurement sets presented in Figs~\ref{fig:beads},~\ref{fig:brain} are summarised in Table~\ref{tab:enhancements}.

\begin{table}[b]
    \caption{\textbf{Key characteristics of the experimental data presented in Figs~\ref{fig:beads},\,\ref{fig:brain}.}
    }
    \centering
    \begin{tabularx}{\columnwidth}{m{30mm} X X X X}
        \toprule
        \textbf{Correction} & \textbf{none} & \textbf{DPP} & \textbf{SLM}\\
        \midrule
        subfigure & (a) & (b) & (c)\\
        display segments &  & 63 & $950^2$\\
        F"~SHARP variant &  & 2-BI & 1-B\\
        $\eta$, beads & 1 & 9 & 14\\
        $\langle\sigma_\eta / \eta\rangle$, beads &  & $\pm 50\,\%$ & $\pm 40\,\%$ \\
        $\eta$, brain cells & 1 & 5 & 8\\
        \bottomrule
    \end{tabularx}
    \label{tab:enhancements}
\end{table}


\section{Discussion}
\label{sec:discussion}

Let us now assess the correction performance of our F"~SHARP variants with respect to several decisive properties, such as optical stability, dependence on sample characteristics, correction speed, and others.

\textit{Optical stability.}
Generally, the performance of an F"~SHARP"=type technique drops quickly and substantially if the relative alignment between static reference and scanned probe beam is suboptimal.
Our Michelson layout keeps the distances which are not common-path between the two beams to a minimum (in our case, about 4~cm round-trip) and is hence far stabler than a (typically much larger) Mach-Zehnder interferometer.
In our experiments, F"~SHARP performance was observed to be stable over a couple of weeks (possibly longer) without requiring realignment.

\textit{DPP vs LCoS-SLM.}
Generally, we find that the SLM yields a higher AO correction performance than the DPP, as expected for several reasons.
First, the (around $7{\times} 10^5$) SLM pixels imaged into our objective pupil provide far more than sufficient degrees of freedom for displaying a correction pattern constructed from 20$\times$20 measured modes, whereas the restriction to 63 DPP electrodes necessarily constitutes a loss of accuracy.
Second, in contrast to the SLM where arbitrary phase shifts can be assigned to each pixel with high precision, the DPP faces mechanical limitations posed by its elastic membrane.
Third, as discussed later, the steps of phase unwrapping and Zernike decomposition (currently) needed for DPP F"~SHARP can decrease the AO performance.
Nevertheless, as we have seen, in the regimes of turbidity studied in this work the DPP can deliver an AO performance of similar quality as the SLM.

\textit{Enhancement spread.}
The relative signal enhancement was found experimentally to vary by about $40\,\%$ (SLM) to $50\,\%$ (DPP), depending on the exact choice of target pixel around a fluorescent bead. 
Thus, if a correction run does not immediately yield a satisfactory result, it can be worthwhile to try again with a nearby target pixel.

\textit{Role of turbidity.}
In regimes of moderate turbidity, as introduced by the scatterer slide (Fig.\,\ref{fig:beads}) or about two hundred microns of brain tissue (Fig.\,\ref{fig:brain}), our experiments showed an aberration correction performance of the DPP slightly lower, but on the same order of magnitude as when using the SLM.
In regimes of very high turbidity, beyond the display capabilities of the DPP, its AO performance drops compared to the SLM. 

\textit{Role of the sample structure.}
Our simulations and experience from experiments confirm that generally, F"~SHARP convergence is faster and correction performance is better for sparse, structured (e.g., beads) than for unstructured (e.g., layer) samples.
This behaviour does not come unexpectedly: in contrast to a point sample, for a homogeneous layer sample initially the sum of signal contributions from many speckles in the focal plane is recorded, each with its individual response to the applied test modes.
It thus takes some iterations until the algorithm ultimately converges onto the strongest speckle.

\textit{Correction speed.}
In our experiment -- currently not highly optimised for speed (data handling, computations, mirror flyback, etc.) -- performing a full F"~SHARP correction typically takes some seconds. 
Let us estimate the potential speed after some experimental refinement.
For correct actuation of the $N_\text S = 63$ segments of the DPP, we need to sample (at least) $N_\text S$ measurement points in the objective pupil using the piezo mirror.
Let us conservatively assume a signal dwell time of $t_\text d = 0.5$~ms per piezo step and $N_\text P = 3$ phase steps per iteration. 
Considering $N_\text I = 3$ iterations and a switching time (including safety margin) of $t_\text s = 100$~ms for the DPP, we arrive at a minimum total correction time $t_\text c =  N_\text I \, (N_\text P \, N_\text S \, t_\text d + t_\text s) \sim 500$~ms.
Hence, achieving F-SHARP AO corrections with the DPP in under a second seems well within reach.
This can be contrasted to the time needed for obtaining a DPP correction using a `traditional' sensorless AO method such as the $3N$-algorithm~\cite{Booth2002_pnas}.
Assuming a case where this algorithm converges, we get $t_\text c =  3 \, N_\text I \, N_\text S \, (t_\text d + t_\text s) \sim 57$~s, clearly highlighting the advantage of the F-SHARP procedure.%

\subsection*{Identified challenges}

\textit{Phase unwrapping.}
We have found in our experiments and simulations that phase unwrapping (as necessary when working with a refractive wavefront modulator such as the DPP) is a critical step which can, if accuracy is insufficient, seriously disturb the AO performance.
Especially when noise is present, standard unwrapping routines often fail to deliver satisfactory results.
However, we have made good experience using more advanced unwrapping methods based on network programming~\cite{Costantini:1998} or on the transport of intensity equation~\cite{Zhao2018}.

\textit{Zernike decomposition.}
We know from numerical simulations that an intermediate change from pixel"=by"=pixel to Zernike basis (variant 2-BI), as currently required in our experiments [Figs~\ref{fig:beads}\,(b), \ref{fig:brain}\,(b)] for technical reasons, is in general detrimental for the F"~SHARP performance. 
The reason is that the Zernike fitting step (especially in combination with imperfect phase unwrapping) leads to a higher percentage of DPP electrodes clipping at the voltage limit, and hence a higher discrepancy between target and displayed phase mask.
We therefore expect that our experimental results could be improved even further by implementing a DPP driver that does not require intermediate Zernike fitting.

\textit{Device limitations.}
The present DPP model features 63 electrodes, which allow for a reasonable F"~SHARP correction performance in moderately turbid regimes.  
To correct for scatterers of a greater thickness, $L$, the DPP would require larger stroke.
To correct for higher spatial frequencies (e.g., larger $\sigma$), the DPP would additionally require a finer resolution, i.e., a larger number of actuator electrodes.


\section{Summary \& Outlook}
\label{sec:summaryoutlook}

To conclude, we have demonstrated scatter correction for multi-photon microscopy using a transmissive, refractive wavefront modulator, the optofluidic deformable phase plate (DPP). 
We have developed a sensorless adaptive optics approach based on a compact and stable Michelson interferometer in combination with a robust, modified F"=SHARP"=type algorithm, which enables fast correction even when using speed-limited wavefront modulators such as the DPP.
In numerical simulations and experiments we have investigated the AO performance with respect to important experimental parameters such as the number of algorithm iterations, sample type, and turbidity regime, and have benchmarked the DPP against an LCoS-SLM.
We have shown that our DPP AO scheme yields significant image improvements for two common microscopy samples suffering from light scattering, fluorescent beads and GFP-labelled brain cells, and have observed a correction performance which is lower, but altogether comparable to an LCoS-SLM.

At this point, let us emphasise again that the DPP itself is still a young development which will surely see further technological improvements in the near future, but also recall the primary aim of this work: 
not to prove superiority of the DPP over an LCoS-SLM, but to demonstrate that F"~SHARP adaptive optics using a DPP \textit{is} feasible, and hence can present an alternative for applications where employing a reflective/""diffractive wavefront modulator is inconvenient or even impossible.

\bigskip

We see great potential for the DPP in combination with our F"~SHARP variant as an add-on AO system which can easily be fitted into pre-existing (e.g., commercial) imaging systems, where a reflective device typically can be accommodated only with severe difficulty.
Near-term directions for future research include implementing a DPP driver which does not require a basis change of the target phase mask into Zernike modes.
Ideally, the DPP influence matrix (which relates the target phase mask to the optimal individual electrode voltages) should be constructed directly in the basis of mechanical eigenmodes of the DPP polymeric membrane. 
This is expected to lead to a lower percentage of electrodes clipping at the voltage limit and hence to better -- and possibly faster converging -- AO corrections.
Longer-term research goals include the development of next-generation DPP devices featuring a higher number of electrodes (i.e., shapeable modes) and/or larger stroke, which would allow to address AO regimes of even higher turbidity, or devices with faster settling times.

\FloatBarrier

\section{Materials \& Methods}
\label{sec:materialsmethods}

\textit{Laser source.}
For our experiments, we used a mode-locked Ti:sapphire laser (MaiTai~DeepSee, MKS~Spectra-Physics, Andover, MA, USA) with about 100~fs pulse duration, a centre wavelength $\lambda_l$ tunable from 700 to 1100~nm, and a maximum time-averaged power of about 200~mW in the sample plane.
In the experiments presented in this work, we set our laser centre wavelength to $\lambda_l = 800$~nm for the bead images and to $\lambda_l = 900$~nm for the glia cell images.

\textit{Imaging system.} 
For general aspects of our TPEF microscopy setup, see Refs~\cite{may2021fast,may2021simultaneous}.
In brief, our laser beam is adjusted in intensity and diameter, passed through the optical setup sketched in Fig.\,\ref{fig:setup}, through a dichroic mirror, and focused by the objective lens (XLUMPLFLN20XW, $\text{NA} = 1$, water immersion, Olympus~Corp., Tokyo, Japan) into the sample plane. 
The epi-TPEF signal is collected by the objective, reflected by the dichroic mirror and focused onto a photomultiplier tube (H10769A-40, Hamamatsu Photonics~K.K., Hamamatsu, Japan) operated in photon counting mode.

\textit{DPP mounting.}
The electromechanical structure of the current DPP is designed such that
it performs virtually identically in vertical and horizontal orientation, i.e., gravitational sag is uncritical~\cite{Rajaeipour:2021}. 
Nevertheless, for our measurements the DPP was mounted in vertical orientation inside a 30-mm cage system, so any potentially detrimental gravity effects would have been maximal.

\textit{Vibration management.} 
Our experiment is set up on a standard optical table with pneumatic vibration isolation. 
We used standard optomechanics and did not take higher vibration prevention efforts than usual for optics experiments.

\textit{Mouse brain samples.}
The brain cells shown in this work were resting CX3Cr-1\textsuperscript{GFP} microglia in coronal slices of mouse brain fixed by perfusion (cf.~Refs~\cite{may2021fast,Sohmen:2022} for details), placed on a microscopy slide, embedded in mounting medium, and covered by a glass slip.
While the fresh brain slices were initially cut to a thickness of around 0.6~mm, until the time of recording the images shown in Fig.\,\ref{fig:brain} (some weeks later) the slices were -- presumably by shrinkage due to evaporation of water -- reduced to a maximum thickness of around 0.2~mm, which is the depth we state in the main text. 
However, this reduction of thickness by a factor of around three might explain why the scattering seems severe for a depth of `only' 0.2~mm.

\textit{Optional scatterer slide.}
The scatterer slide was prepared through rough manual application of transparent nail polish to a glass slide and letting dry.
It was subsequently placed in the excitation path in conjugate position to a plane in the focusing cone of the objective lens (cf.~Fig.\,\ref{fig:setup}).
Being located before the dichroic mirror, the aberrations by the nail polish slide affect only the excitation light, \textit{not} the emitted fluorescence light. 
However, since for signal recording we collect all photons emitted into our objective's numerical aperture (without, e.g., re-imaging onto a pinhole as in confocal microscopy), this already captures the critical difficulty for typical multi-photon imaging applications.

\textit{Signal enhancement.}
We calculate the enhancement factor for subfigure (x) relative to subfigure (a) in the image set as 
$\eta_x = \max_m{(I_x)} / \max_m{(I_a)}$, 
where $\max_m{(\ldots)}$ is the mean over the respective brightest $m = 9$ pixels of the image.

\textit{Relative spread of enhancement.}
To assess the variation of the AO algorithm performance with the choice of target pixel, we investigated 5 different beads of the same kind as shown in Fig.\,\ref{fig:beads}.
For each bead, 3 pairs of measurements were taken.
Each pair of measurements consisted of (i) picking a \textit{new} target pixel in the region where the bead is suspected, (ii) running F"~SHARP with the SLM (variant 1-B), and then (iii) anew with the DPP (variant 2-BI).
Note that by interleaving measurements with SLM and DPP we have aimed to mitigate systematic errors due to, e.g., fluorophore bleaching.
From the respective 3 runs per bead we calculate the relative standard deviation of enhancement, $\sigma_\eta / \eta$, for both variants.
Finally, we compute the corresponding mean values over the five beads, $\langle\sigma_\eta / \eta\rangle$.

~


\section*{Conflict of Interest Statement}

The Innsbruck team declares that its research was conducted in the absence of any commercial or financial relationships that could be construed as a potential conflict of interest.
Members of the Freiburg team are with the commercial company Phaseform GmbH.

\section*{Author Contributions}

The Innsbruck team conceived, assembled, and conducted the experiments. 
M.S.~and J.D.M.B.~took the experimental data.
M.S.~and A.J.~performed the numerical simulations.
M.S.~wrote the manuscript with input from all other authors. 
The Freiburg team provided the DPP device, practical advice on how to operate it, the DPP driver software, and valuable input for Section~\ref{sec:dpp}.

\section*{Funding}

The Innsbruck team acknowledges funding from the Austrian Science Fund (FWF) under the grant numbers~P32146-N36, M~3060-NBL, and I3984.

\section*{Acknowledgements}

We would like to thank Kai K.~Kummer and Jeiny Luna Choconta, Institute for Physiology, Medical University of Innsbruck, for preparing the brain sample.

\section*{Data Availability Statement}

The data sets presented in this work are willingly available upon reasonable request.

\bibliography{references.bib}

\end{document}